\input harvmac
\def\half{{1 \over 2}}

\def\>{{\rangle}}
\def\<{{\langle}}

\def\p{{\partial}}
\def\pt{{\partial_\tau}}
\def\s{{\sigma}}

\def\l{{\lambda}}
\def\lb{{\bar\lambda}}

\def\a {{\alpha}}
\def\b {{\beta}}
\def\g {{\gamma}}
\def\ad {{\dot \a}}
\def\bd {{\dot \b}}
\def\gd {{\dot \g}}

\def\G {{\Gamma}}
\def\d {{\delta}}

\def\e {{\epsilon}}

\def\k {{\kappa}}
\def\kb {{\bar\kappa}}
\def\t {{\theta}}

\def \t {{\theta}}
\def \T {{\Theta}}
\def \tb {{\bar\theta}}
\def \Tb {{\bar\Theta}}

\Title{\vbox{\hbox{IFT-P.033/97}}}
{\vbox{\centerline{\bf Extra Dimensions in Superstring Theory}}}
\bigskip\centerline{Nathan Berkovits}
\bigskip\centerline{Instituto
de F\'{\i}sica Te\'orica, Univ. Estadual Paulista}
\centerline{Rua Pamplona 145, S\~ao Paulo, SP 01405-900, BRASIL}
\bigskip\centerline{e-mail: nberkovi@ift.unesp.br}
\vskip .2in
    It was earlier shown that an SO(9,1) $\theta^\a$ spinor variable 
can be constructed from RNS matter and ghost fields. $\theta^\a$ has 
a bosonic worldsheet super-partner $\lambda^\a$ which plays the role 
of a twistor variable, satisfying  
$\lambda\Gamma^\mu\lambda = \partial x^\mu +i\theta\Gamma^\mu \partial\theta$.
For Type IIA superstrings, the left-moving $[\theta_L^\a,\lambda_L^\a]$ 
and right-moving $[\theta_{R\a},\lambda_{R\a}]$ can be combined into 
32-component SO(10,1) spinors $[\theta^A,\lambda^A]$. 

    This suggests that $\lambda^A \Gamma^{11}_{AB}\lambda^B=
2\lambda_L^\a \lambda_{R\a}$ can be interpreted as momentum in the eleventh 
direction. Evidence for this interpretation comes from the zero-momentum
vertex operators of the Type IIA superstring and from consideration of 
$D_0$-branes. As in the work of Bars, one finds an SO(10,2) structure for the 
Type IIA superstring and an SO(9,1) x SO(2,1) structure for the Type IIB 
superstring. 

\Date{April 1997}
\newsec {Introduction}

There is accumulating evidence that ten-dimensional superstring theory is
related to a theory in eleven dimensions.\ref\dual
{C. Hull and P.K. Townsend, Nucl. Phys. B438 (1995) 109.}\ref\witten
{E. Witten, Nucl. Phys. B443 (1995) 85.}\ref\polch{J. Polchinski,
Phys. Rev. Lett. 75 (1995) 4724.}
Since most information about this
eleven-dimensional theory comes from compactification or from
low-energy analysis
of supergravity, little is known about its fundamental nature. Most proposals
for understanding the extra dimension introduce a new fundamental object,
the
supermembrane, whose 
double-dimensional reduction gives the Type IIA superstring.\ref\supermem
{M.J. Duff, P.S. Howe, T. Inami and K.S. Stelle, 
Phys. Lett. 191B (1987) 70\semi
P.K. Townsend, Phys. Lett. B350 (1995) 184\semi
P.K. Townsend, ``p-Brane Democracy'', hep-th 9507048\semi
M.J. Duff, ``Supermembranes'', hep-th 9611203.}

In this paper, it will be proposed that the extra dimension can be 
obtained from the usual superstring theory without introducing new
fundamental objects. (Although D-branes are present in the non-perturbative
superstring spectrum, they are not fundamental objects in the sense that
superstrings do not come from their dimensional reduction.) Since superstring
theory only contains ten $x$'s, it is natural to ask where the extra
dimension comes from. 

In the RNS description of superstrings, one has
super-worldsheet ghosts, $[b,c]$ and $[\b,\g]$,
which are crucial for constructing
Ramond vertex operators and spacetime-supersymmetry generators. In this
paper, it will be proposed that the bosonic
variable for the extra dimension comes
from a particular Ramond-Ramond combination of RNS matter and ghost fields. 
The appropriate Ramond-Ramond combination of fields is found by
constructing twistor-like variables for the superstring. These twistor-like
variables first appeared in the GS description of the superstring.\ref\tw
{D.P. Sorokin, V.I. Tkach, D. V. Volkov and A.A. Zheltukhin, Phys. Lett. B216
(1989) 302.}

The standard GS description of the superstring contains fermionic Siegel
symmetries rather than worldsheet supersymmetries, which has prevented a
successful quantization except in light-cone
gauge. However, there exists a modified GS superstring
which can be quantized (although not with manifest SO(9,1) invariance) 
and which contains bosonic spinor variables, $\l^\a$ and $\bar\l^\a$, in
addition to the usual GS variables, $x^\mu$ and $\theta^\a$. 
\ref\tonin{M. Tonin, Phys. Lett. B266 (1991) 312\semi
E.A. Ivanov and A.A. Kapustnikov, Phys. Lett. B267 (1991) 175.}
\ref\het{N. Berkovits, Nucl. Phys. B379 (1992) 96.}
These bosonic
spinors are not independent fields, but satisfy the twistor-like 
constraint\tw \foot{The
unusual factor of ${i\over 2}$ is used 
so that $\{q_\a,q_\b\}=P_\mu \G^\mu_{\a\b}$
rather than $2P_\mu \G^\mu_{\a\b}$.}
\eqn\one{\lambda^\a \Gamma^\mu_{\a\b} \bar\lambda^\b 
=\p x^\mu +{i\over 2}\t^\a \G^\mu_{\a\b} \p\t^\b, }
as well as the pure spinor constraint
$\lambda^\a \Gamma^\mu_{\a\b}\lambda^\b =
\bar\lambda^\a \Gamma^\mu_{\a\b} \bar\lambda^\b =0.$\tonin

In this twistor version of the GS superstring, two of the eight Siegel
symmetries are replaced with N=2 worldsheet supersymmetries. 
(Although there is also a twistor version of the GS superstring where all
eight Siegel symmetries are replaced with worldsheet supersymmetries,\ref
\howe{F. Delduc, A. Galperin, P. Howe and E. Sokatchev, Phys. Rev. D47
(1993) 578.}
this N=8 twistor version of the GS superstring has not yet been quantized.)
Under the N=2 worldsheet supersymmetry transformations, the
$\t^\a, \l^\a, \bar\l^\a$, and $x^\mu$ fields transform as
components of the N=2 superfields
\eqn\dft{\Theta^\a=\theta^\a +\k \l^\a +\bar\k\bar\l^\a +\k\bar\k h^a,}
$$X^\mu =x^\mu +i\k m^\mu +i\bar\k \bar m^\mu +\k\bar\k n^\mu,$$
satisfying the twistor and pure spinor constraints:
\eqn\two{{i\over 2} \T^\a\Gamma^\mu_{\a\b} D\T^\b =D X^\mu,\quad
{i\over 2}\T^\a \Gamma^\mu_{\a\b} \bar D\T^\b =\bar D X^\mu,}
where $D= d/d\k +{i\over 2}\bar\k \p_z$,
$\bar D= d/d\bar\k +{i\over 2}\k \p_z$,
and $f^\a, m^\mu, n^\mu$ are auxiliary fields. 

For the Type IIA superstring, the left-moving $\T^\a_L$
carry SO(9,1) Weyl spinor indices while the right-moving 
$\T_{R\a}$ carry SO(9,1) anti-Weyl spinor indices.
This allows them to be combined into a 32-component SO(10,1) spinor 
superfield
$\T^A$.\foot{ In this paper, Greek letters are SO(9,1) indices,
capitalized Latin letters
are SO(10,1) or SO(10,2) indices, and uncapitalized
Latin letters are SO(2,1) indices. 
Letters from the first half of the alphabet denote
spinor indices and letters from the second half of the alphabet denote
vector indices. 
The SO(9,1) vector
indices will take the values 0 ... 9,
the SO(10,1) vector indices will take the values 0, ..., 9, 11, the 
SO(10,2) vector indices will take the values 0, ..., 9, 11, 12, and
the SO(2,1) vector indices will take the values 0,1,2.
The flat metric is $\eta_{00}=\eta_{12\,12}=1$ and $\eta_{MM}=-1$ for
$M=1 ... 9, 11.$
SO(9,1) $\Gamma^\mu$ matrices are 16 $\times$ 16 and satisfy
$\G^{(\mu}_{\a\b} \G^{\nu)\,\b\g}=2 \eta^{\mu\nu} \d_\a^\g$. SO(10,1) and
SO(10,2) $\G^M$ matrices are 32 $\times$ 32 and satisfy 
$\G^{(M}_{AB} \G^{N)\,BC}=2 \eta^{MN} \d_A^C$. For SO(9,1) and SO(10,2), these
$\G$ matrices
are related to the usual 32$\times$32 and 64$\times$64
$\g$ matrices by multiplication with $\g^0$ and by taking the upper diagonal
quadrant. For SO(10,1), they are related to the usual 32$\times$32
$\g$ matrix by multiplication with $\g^0$. In other words,
$\G^M_{AB}=(\g^0\g^M)_A^B$ and
$\G^{M\,AB}=(\g^M\g^0)_A^B$.
The explicit representation for $\G^M$ will be $\G^0_{AB}=\d_{AB}$,
$\G^M_{AB}=\sigma_3 \times \G^M_{\a\b}$ for $M=1 ... 9$,
$\G^{11}_{AB}=\sigma_1 \times 1_{16}$, and
$\G^{12}_{AB}=i\sigma_2 \times 1_{16}$ where $\sigma^i$ are the
Pauli matrices and $1_{16}$ is the $16\times 16$ identity matrix.
Note that $\G^M_{AB}=\G^M_{BA}$ except when $M=12$, and
$\G^{12}_{AB} =-\G^{12}_{BA}$.}

The natural higher-dimensional generalization of the twistor constraint is 
\eqn\thr{\l^A \Gamma^M_{AB} \bar\l^B = P^M }
where $P^{11}$ is defined by this constraint, i.e.
\eqn\hgh{P^{11}= \l^A \Gamma^{11}_{AB} \bar\l^B 
= \l^\a_L \bar\l_{R\a}+\l_{R\a}\bar\l^\a_L.}
In fact, one can also interpret $\l^A$ and $\bar\l^B$ as SO(10,2) 
Majorana Weyl and Majorana anti-Weyl spinors, in
which case $P^{12}=
\l^\a_L \bar\l_{R\a}-\l_{R\a}\bar\l^\a_L.$ 

To translate this into RNS language (where covariant 
quantization is known), one needs to find the combination of RNS matter and
ghost fields which corresponds to $\l^\a$ and $\bar\l^\a$.
Fortunately, the dictionary between RNS fields and twistor-GS fields was
found in reference \ref\equal
{N. Berkovits, Nucl. Phys. B420 (1994) 332.}
where it was shown how to explicitly construct
$\l^\a$ and $\bar\l^\a$ in terms of the RNS matter and ghost fields.
It was also shown in this reference that the fermionic
N=2 superconformal generators
of the twistor-GS superstring are mapped in RNS language into the RNS BRST
current and the $b$ ghost.
So the twistor variables, $\l^\a$ and $\bar\l^\a$, are obtained in RNS
language by anticommuting the $\t^\a$ variable with the BRST charge and
with the $b$ ghost.\foot{Recently, 
Dimitri Polyakov has expressed related ideas.\ref\pol{D. Polyakov,
Nucl. Phys. B485 (1997) 128.}\ref\poltwo{D. Polyakov, ``S Duality as an
Open String Gauge Symmetry'', hep-th 9703008.}
However, there are some crucial differences between our approaches. Firstly,
he defines his twistor variable, $\l^\a$, as the anticommutator of $\t^\a$
with the N=1 RNS superconformal generator. Therefore, his definition of
$\l^\a$ is not GSO-projected, i.e. it has square-root cuts with the
spacetime-supersymmetry generators. Secondly, he only considers left-moving
twistor variables so there is no analog of $\l^\a_L \l_{R\a}$. Although
he claims in \poltwo
that the anticommutator of left-moving spacetime-supersymmetry
generators in the $+1/2$ picture 
has a five-form central charge proportional to
$\Gamma^{\mu_1 ... \mu_5} \psi_{\mu_1} ... \psi_{\mu_5}$, his computation
appears to be incorrect. My calculation of the five-form term in this
anticommutator gives something proportional to 
$\G_\nu \G^{\mu_1 ...\mu_5}\G^\nu$, which vanishes in ten dimensions.}

In section 2, the dictionary between the RNS and twistor-GS
variables is reviewed. The twistor-like variables, $\l^A$ and $\bar\l^A$,
are explicitly constructed in terms
of RNS matter and ghost fields. 

In section 3, 
the identification of $P^{11}$ with ${1\over{32}}(\l_L^\a \bar\l_{R\a}+
\l_{R\a} \bar\l_L^\a)$ is justified by analyzing zero-momentum
vertex operators for 
massless states of the Type IIA superstring, which correspond to the
zero-momentum spectrum of
D=11 supergravity.  
For NS-NS states, these vertex operators are well-known, but for R-R
states, these vertex operators are new and are constructed using
the R-R sector of closed superstring field theory.\ref\sft
{N. Berkovits, Phys. Lett. B388 (1996) 743.} Although it is often
stated 
that the R-R vertex operator vanishes at zero momentum, this
is not completely true. It will be shown that the zero-momentum
R-R vertex operator is BRST-equivalent to an operator of ghost-number
$(1+2n,1-2n)$ where $n$ is arbitrarily large. 
This allows the
construction of a field theory action for the massless R-R string 
fields\sft\ref\sftwo{N. Berkovits, Phys. Lett. B395 (1997) 28.}
and
also implies that all vertex operators with finite ghost-number must
decouple from the zero-momentum R-R vertex operators.
The structure of the zero-momentum R-R vertex operator suggests that
the general $p$-brane R-R charge can be constructed from RNS variables
in a manner similar to the zero-brane charge ${1\over{32}}
\oint d\sigma \l\G^{11}\lb$.

Actually, sigma model arguments imply that
it is the R-R gauge field times the exponential of the dilaton,
$e^{\phi}$, which couples
to these zero-momentum R-R vertex operators
so
$P_{11}$ should really be identified with
${1\over{32}} e^{\phi} (\l \Gamma_{11} \bar\l)$.
In section 4, $D_0$-branes are shown to be massless in eleven dimensions if
the zero-brane charge, $P_{11}$, is identified with
${1\over{32}} e^{\phi} (\l \Gamma_{11} \bar\l)$. 

In section 5, these techniques are generalized to the Type IIB superstring
where the SO(10,2) structure is replaced by an SO(9,1)$\times$ 
SO(2,1) structure.
These SO(10,2) and SO(9,1)$\times$ SO(2,1) 
structures were also found by Bars in
reference \ref\Bars{I. Bars, Phys. Rev. D55 (1997) 2373\semi
I. Bars, ``Algebraic Structure of S Theory'', talk at Strings '96, hep-th
9608061.}.

Finally, in section 6, some connections are made with other proposals to 
understand the eleventh dimension.

\newsec{Construction of twistor variables}

\subsec{Review of GS - RNS dictionary}

In the RNS description of the D=10 superstring, the spacetime supersymmetry
generator 
\eqn\su{q_\a =\oint dz ~e^{-\phi/2} \Sigma_\a}
satisfies the algebra $\{q_\a~, ~q_\b\}=\oint dz ~ e^{-\phi} \psi_\mu
\G^\mu_{\a\b}$ where $\Sigma_\a$ is the Ramond spin field of weight $5/8$,
and
the $\beta$ and $\g$ worldsheet ghosts have been
fermionized as $\b=\p\xi e^{-\phi}$ and $\g=\eta e^{\phi}$.
Although $e^{-\phi}\psi_\mu$ is related by picture-changing to the 
momentum operator $\p x^\mu$, this is not good enough for manifest
spacetime supersymmetry since picture-changing is only an on-shell
operation. 

One therefore needs to introduce a second spacetime-supersymmetry generator
\eqn\suba{\bar q_\a =\oint dz ( e^{\phi/2} \Sigma^\b \p x_\mu \G^\mu_{\a\b} +
b \eta e^{3\phi/2} \Sigma_\a )}
which is BRST invariant and is related to $q_\a$ by picture-changing.
(Note that $\bar q_\a$ is Majorana-Weyl and is not the complex conjugate
of $q_\a$.) It is easy to check that $\{q_\a ~,~\bar q_\b\}$ =
$\G^\mu_{\a\b}\oint dz ~ \p x_\mu$ as desired.

However, since  
$\{q_\a ~,~ q_\b\}$ does not vanish, this is not a standard N=2 D=10
supersymmetry algebra.\foot{For compactifications 
to four dimensions which preserve N=1 D=4 supersymmetry, one can choose
the two chiral N=1 D=4 supersymmetry generators 
to come from $q_\a$ and the two anti-chiral 
supersymmetry generators to come from $\bar q_\a$.
In this case, $\{q_\a,q_\b\}=0$ which 
allows a formulation of the superstring with manifest SO(3,1)
super-Poincar\'e invariance.\ref\calabi{N. Berkovits, Nucl. Phys. B431
(1994) 258.}}
Nevertheless, it will be useful to define two spinor variables, $\t^\a$
and $\tb^\a$, which satisfy the anti-commutation relations
$\{q_\a~,~ \t^\b\}=
\{\bar q_\a ~,~\bar\t^\b\}=\d_\a^\b$. These are easily found to be
\eqn\thetadef{\t^\a = e^{\phi/2} \Sigma^\a,\quad
\bar\t^\a = c \xi e^{-3\phi/2} \Sigma^\a.}
Note that $\tb^\a$ involves the $\xi$ zero mode, which is necessary
for preserving manifest spacetime supersymmetry.

As shown in reference \equal, the N=2 worldsheet superconformal generators
in the twistor-GS formalism are mapped into the following RNS expressions:
$$T= \half \p x_\mu \p x^\mu +{i\over 2} \psi_\mu \p\psi^\mu +2i b\p c 
-i c\p b
+i\eta \p\xi +\half \p \phi \p\phi +\p^2 \phi -{i\over 2}\p (bc +\xi\eta),$$
$$G =\eta e^{\phi} \psi^\mu \p x_\mu + i\eta\p\eta e^{2\phi} b +\p (i c\xi\eta
+\p c)$$
$$+c 
(\half \p x_\mu \p x^\mu +{i\over 2} \psi_\mu \p\psi^\mu +i b\p c 
+ i\eta \p\xi +\half \p \phi \p\phi +\p^2 \phi),$$
\eqn\superc{\bar G =b,}
$$J=cb +\eta\xi.$$
These generate a $c=6$ N=2 superconformal algebra and, after redefining
$T \to T -{i\over 2}\p J$, form a set of twisted N=2 generators
whose $T$ is the standard
RNS stress-energy tensor, $G$ is the BRST current, $\bar G$ is the $b$ ghost,
and $J$ is the RNS ghost-number current. 
(Although the RNS ghost-number charge
is
usually defined by $\oint dz~(cb-i\p\phi)$, this agrees with
$\oint dz~J$ at zero picture, i.e. when $\oint dz~(\eta\xi+i\p\phi)=0$.)

It is natural to ask how the $x^\mu$, $\t^\a$ and $\tb^\a$ variables
transform under commutation with the above generators. One finds that
$\{\t^\a~,~ \oint dz~\bar G\}=0$ so $\t^\a$ is the lowest component of an N=2
chiral superfield, 
$\T^\a =\t^\a +\k\l^\a +{i\over 2}\k\kb\p\t^\a$ where
\eqn\lam{\l^\a =\{ \oint dz~
G~,~\t^\a \}= \eta e^{3\phi/2}\p x^\mu \Sigma_\b \G_\mu^{\a\b}
+ b \eta \p\eta e^{5\phi/2}\Sigma^\a +c \p (e^{\phi/2} \Sigma^\a),}
$D=d/d\k +{i\over 2}\kb\p_z$ and
$\bar D=d/d\kb +{i\over 2}\k\p_z$.
Similarly,
$\{\tb^\a~,~ \oint dz ~G\}=0$ so $\tb^\a$ is the lowest component of an N=2
anti-chiral superfield, 
$\Tb^\a =\tb^\a +\kb\lb^\a -{i\over 2}\k\kb\p\tb^\a$ where 
\eqn\lamb{\lb^\a =\{ \oint dz~
\bar G~,~\tb^\a \}= \xi e^{-3\phi/2} \Sigma^\a.}
Finally, $[ x^\mu~,~\oint dz~\bar G]$=0 implies that $x^\mu$ is the lowest
component of an N=2 chiral superfield
$X^\mu =x^\mu +\kappa \chi^\mu +{i\over 2}\k\kb\p x^\mu$ where
$\chi^\mu =[\oint dz ~G~,~x^\mu]= \eta e^{\phi} \psi^\mu$.

Using the usual RNS OPE's, these superfields can be shown to 
satisfy the constaint
\eqn\const{i(D\T^\a) \Tb^\b = \G_\mu^{\a\b} D X^\mu,}
which implies the twistor-like condition\ref\cov{N. Berkovits, Phys. Lett. 
B300 (1993) 53.}\equal
\eqn\asdf{\l^\a\lb^\b = \G_\mu^{\a\b} \p x^\mu -i(\p\t^\a)\tb^\b .}
After defining $\hat X^\mu=X^\mu -{i\over{32}}
\T\G^\mu\Tb$, the constraint of \const can be rewritten as
\eqn\consttwo{{i\over 2}(D\T^\a) \Tb^\b = \G_\mu^{\a\b} D\hat X^\mu,\quad 
{i\over 2}(\bar D\Tb^\a) \T^\b = \G_\mu^{\a\b} \bar D\hat X^\mu,}
which resembles the twistor-GS constraint of \two.

However, the RNS constraint of \asdf
has 256 components, rather than the 10 components
of \one, and there is no pure spinor constraint.
Furthermore, there are two spinor variables, $\t^\a$ and $\tb^\a$,
in the RNS approach while there is only one spinor variable in the twistor-GS 
approach. These differences come from the fact that the twistor-GS
superstring has six Siegel symmetries in addition to the two worldsheet
supersymmetries. The equivalence between the RNS and twistor-GS formalisms
has only been proven after gauge-fixing these six Siegel symmetries by
setting six of the components of $\T^\a_{GS}$ to zero. In this
non-covariant gauge, the remaining ten
components of $\T^\a_{GS}$ split into two pure spinors, one of which
is an N=2 chiral superfield identified with $\T_{RNS}^\a$, and the other
is an N=2 anti-chiral superfield which is identified with $\Tb_{RNS}^\a$.\het
It is then straightforward to prove the equivalence of the two 
formalisms.\equal

For the rest of this paper, only the RNS formalism will be discussed.

\subsec{Construction of the extra dimension}

For the Type IIA superstring, one can construct
left and right-moving superfields, $(\T^\a_L, \Tb^\a_L)$ and
$(\T_{R\a}, \Tb_{R\a})$, which carry Weyl and anti-Weyl SO(9,1) spinor
indices. They can therefore be combined into 32-component superfields
$(\T^A,\Tb^A)$ 
which transform
as SO(10,1) Majorana spinors. In fact, one can also interpret them as
32-component
SO(10,2) spinors where $\T^A$ transforms as a Majorana Weyl spinor and
$\Tb^A$ transforms as a Majorana anti-Weyl spinor. With this choice of
SO(10,2)
chirality, $\T^A \G^M_{AB} \Tb^B$ transforms as an SO(10,2) vector.

So an obvious generalization of \const is
\eqn\obv{{i\over{16}}(D\T^A) \G^M_{AB} \Tb^B = D X^M,}
where $D=d/d\k +i\kb \p_\tau$ and 
$\bar D=d/d\kb +i\k \pt$, $\pt= \half(d/dz_L +d/dz_R)=\half(\p_L+\p_R)$,
$\p_\sigma=\half(\p_L-\p_R)$,
$\T^A=\t^A +\k\l^A +i\k\kb \p_\tau \t^A$,
$\bar \T^A=\tb^A +\kb\lb^A -i\k\kb \p_\tau \tb^A$,
and $X^M$ is a chiral N=2 superfield defined by \obv,
i.e. 
\eqn\eleven{\p_\tau (x^{11} + x^{12}) ={1\over{16}}(\l_L^\a \bar\l_{R\a} +
2i(\p_\tau \t_L^\a) \tb_{R\a}),}
$$\p_\tau (x^{11} - x^{12}) ={1\over{16}}(\l_{R\a} \bar\l_L^\a 
+2i(\p_\tau \t_{R\a}) \tb_L^\a ).$$
The N=2 worldsheet supersymmetry generators are now the sum of the
left-moving and right-moving N=2 superconformal generators of \superc.
(This is consistent with the definition of $D$ and $\bar D$ since
$\p_\tau\T_L^\a=\half\p_L\T_L^\a$ and
$\p_\tau\T_R^\a=\half\p_R\T_R^\a$.)
For $M=0$ to 9, $x^M$ is easily seen to be defined in the same
way as in \asdf, i.e.
\eqn\same{\p_\tau x^\mu ={1\over{32}}(\l_L \G^\mu \bar\l_L +
\l_R \G^\mu \bar\l_R +i
(\p_L \t_L)\G^\mu \tb_L
+i(\p_R \t_R)\G^\mu \tb_R).}
Note that the stronger condition,
$i (D\T^A) \Tb^B =\G_M^{AB} D X^M$
cannot be correct since $\l_L^\a \lb_{R\b}$ is not proportional to
$\d^\a_\b$. Also note that the RNS definition of 
\eqn\rnsd{P_M ={1\over{32}}\l^A\G^M_{AB}\lb^B}
differs by a factor of 32 from the GS definition of \thr.

In the following two sections, the above identification of $\pt x^{11}$
will
be justified using arguments based on superstring vertex operators and
on $D_0$-branes.

\newsec{Justification based on Type IIA zero-momentum vertex operators}

\subsec{Zero-momentum NS-NS vertex operators}

The zero-momentum
states of the D=10 Type IIA superstring match the zero-momentum
states of D=11 supergravity.
Under compactification on a circle, the D=11 graviton decomposes into
a D=10 graviton, dilaton, and graviphoton, and the D=11 three-form 
decomposes into a D=10 three-form and two-form.
Since the zero-momentum graviton, $g_{\mu\nu}$, has vertex operator
$\int d^2 z ~\p_\tau x^\mu \p_\tau x^\nu$ (ignoring the
$\p_\sigma x^\mu$ dependence), one might expect the 
vertex operators of the
zero-momentum dilaton and graviphoton, $\phi$ and $A_\mu$, to be related to
$\int d^2 z ~(\p_\tau x^{11})^2$ and
$\int d^2 z ~\p_\tau x^{11}\p_\tau x^\mu$.

Plugging in the definition of \eleven for $\pt x^{11}$ (and ignoring the
$\t$ dependence), one finds 
\eqn\fone{\int d^2 z ~(\p_\tau x^{11})^2={1\over{2^{10}}}
\int d^2 z ~(\l_L^\a\lb_{R\a} +\l_{R\a}\lb^\a_L)^2,}
which is not a simple expression in terms of RNS free fields.
However, consider instead
\eqn\ftwo{\int d^2 z ~[(\p_\tau x^{11})^2 -(\p_\tau x^{12})^2]={1\over{2^8}}
\int d^2 z ~\l_L^\a\lb_{R\a} \l_{R\b}\lb^\b_L.}
Using the identity of \asdf (and ignoring the $\t$ dependence), one finds 
\eqn\fthr{\int d^2 z ~[(\p_\tau x^{11})^2 -(\p_\tau x^{12})^2]
={1\over{16}}\int d^2 z~ [\p_L x^\mu \p_R x_\mu],}
which is proportional to the zero-momentum dilaton vertex operator in
integrated form.\foot{In unintegrated form, there are two physical
zero-momentum dilaton vertex operators, $c_L \p_L x^\mu c_R \p_R x_\mu$ and
$c_L \p_L^2 c_L +(\eta_L e^{\phi_L})\p_L(\eta_L e^{\phi_L})=
\{Q_L, \p_L c_L\}.$\ref\bel{A. Belopolsky and B. Zwiebach, Nucl. Phys. B472
(1996) 109.} The integrated form is obtained by anti-commuting the
unintegrated form with $\int dz_L b_L$ and $\int dz_R b_R$, so the
second type of ``ghost'' dilaton vertex operator decouples in the
absence of worldsheet curvature (worldsheet curvature can mix
$b_L$ with $b_R$). Note that 
$\int d^2 z ~[(\p_\tau x^{11})^2 -(\p_\tau x^{12})^2]$ needs to be
normal-ordered in the presence of worldsheet curvature. It would be
interesting to see if this normal-ordering procedure is somehow related
to the ``ghost'' dilaton vertex operator.}

So with the twistor definition of $\pt x^{11}$ and $\pt x^{12}$, the dilaton
appears to be related to the $(11,11)-(12,12)$ components
of a twelve-dimensional graviton. 
(The (11,11)+(12,12) component appears not to
have a simple string interpretation.)
This suggests that the Type IIA
dilaton measures the volume of the torus which compactifies from 
$10+2$ to $9+1$ dimensions. 

\subsec{Zero-momentum R-R vertex operators}

It is commonly stated that R-R gauge fields decouple from strings
at zero momentum. This statement is based on three arguments: 1)
The standard massless R-R vertex operator vanishes at zero momentum; 2)
There are no coupling terms of the appropriate dimension in the standard GS
sigma model; 3) No perturbative superstring states carry R-R charge.

However, if the above statement were true, it would be impossible to
construct a superstring field theory action in the R-R sector since
there is no Maxwell action without gauge fields. In recent papers \sft\sftwo,
such an action was constructed, and it will now be explained how
superstring field theory solves this problem without violating the
above three arguments.

The superstring field theory action comes from a $<\Phi Q \Phi>$ action
where $\Phi$ is the superstring field and, for the massless Type IIA R-R
sector, $\Phi$  contains infinite copies of four bispinor fields:
$C_{(n)\a}^\b$, $D_{(n)\a\b}$, 
$E^{(n)\a\b}$, 
$F_{(n)\b}^\a$ for $n$=0 to $\infty$. (The infinite copies come from the
dependence of $\Phi$ on the $\beta,\gamma$ zero modes.)
The action for these fields can be found in \sft\sftwo, and it was shown
that all the $C_{(n)\a}^\b$ fields and all but one of the $D_{(n)\a\b}$
and $E_{(n)}^{\a\b}$ fields can be gauged away. The 
equations of motion in this gauge are
\eqn\motion{
F_{(0)\b}^\a =\p_\mu (D_{\g\b}\G^{\mu~\a\g}-E^{\a\g}\G^\mu_{\g\b}),}
$$\G^\mu_{\a\g}\p_\mu F^{\a}_{(0)\b}=
\G_\mu^{\g\b}\p^\mu F^{\a}_{(0)\b}=0, \quad
F_{(n)\b}^\a=0 ~~for ~ n>0.$$
Note that $F^\a_{(0)\b}$ is an auxiliary field which satisfies Bianchi
identities only on-shell.

Although this superstring field theory action was constructed using
``non-minimal'' RNS fields\ref\sieg
{W. Siegel, Int. J. Mod. Phys. A6 (1991) 3997\semi
N. Berkovits, M.T. Hatsuda and W. Siegel, Nucl. Phys. B371 (1991) 434.},
one can analyze the vertex operators for
the gauge fields, $D_{\a\b}$ and $E_{\a\b}$, using the usual minimal
set of RNS fields. For simplicity, these vertex operators will be analyzed
at zero momentum.

Consider the following R-R vertex operator in unintegrated form:
\eqn\vdef{ V_{(0)}^{\a\b}= c_L e^{-3\phi_L/2} \Sigma_L^\a ~
c_R e^{-\phi_R/2} \Sigma_R^\b .}
This operator is naively BRST-trivial since $V_{(0)}^{\a\b}=[Q_L+Q_R~,~
(\p_L c_L) W_{(0)}^{\a\b}]$ where
\eqn\wdef{W_{(0)}^{\a\b}=
 c_L \p_L \xi_L \p_L^2 \xi_L e^{-7\phi_L/2} \Sigma_L^\a ~
c_R e^{-\phi_R/2} \Sigma_R^\b .}
However, since $(b^0_L -b^0_R) \p_L c_L W_{(0)}^{\a\b} \neq 0$,\foot
{I would like to thank Sanjaye Ramgoolam for pointing this out to me.} 
$V_{(0)}$ is not BRST-trivial but is in the same semi-relative
cohomology class as
$[Q_L+Q_R~,~
(\p_R c_R) W_{(0)}^{\a\b}]=V_{(1)}^{\a\b}$ 
where
\eqn\vone{V_{(1)}^{\a\b}=
 c_L \p_L \xi_L \p_L^2 \xi_L e^{-7\phi_L/2} \Sigma_L^\a ~
c_R e^{3\phi_R/2} \eta_R \p_R \eta_R\Sigma_R^\b .}
($b_L^0-b_R^0$ signifies the zero mode of $b_L-b_R$ and semi-relative
cohomology is defined in \ref\csf{B. Zwiebach, Nucl. Phys. B390 (1993) 33.}.)

Similarly, $V_{(1)}^{\a\b}$ is not BRST-trivial, but is in the same cohomology
class as
$$V_{(2)}^{\a\b}=
c_L \p_L \xi_L \p_L^2 \xi_L
\p_L^3 \xi_L \p_L^4 \xi_L
e^{-11\phi_L/2} \Sigma_L^\a ~
c_R e^{7\phi_R/2} 
\eta_R \p_R \eta_R
\p_R^2\eta_R \p_R^3 \eta_R\Sigma_R^\b .$$
This chain continues forever, so $V_{(0)}^{\a\b}$ is in the same BRST
cohomology class as $V_{(n)}^{\a\b}$ for arbitrarily large $n$ where
$V_{(n)}^{\a\b}$ carries ghost number $(1-2n,1+2n)$. (The ghost-number is
defined by commuting with $[J_L,J_R]$ of \superc.)
Also, 
\eqn\vbar{ V_{(0)\a\b}= c_L e^{-\phi_L/2} \Sigma_{L\a} ~
c_R e^{-3\phi_R/2} \Sigma_{R\b}}
is in the same BRST
cohomology class as $V_{(n)\a\b}$ for arbitrarily large $n$ where
$V_{(n)\a\b}$ carries ghost number $(1+2n,1-2n)$. 

Since these vertex operators, $V_{(n)}^{\a\b}$ and $V_{(n)\a\b}$,
have the same BRST structure as the
string field for $D_{(n)\a\b}$ and $E_{(n)}^{\a\b}$ in \sft\sftwo, 
they will be conjectured
to be equivalent. (This is a conjecture since it is not yet known how
to construct a superstring field theory action without introducing 
``non-minimal'' RNS fields.)

So zero-momentum R-R vertex operators are present in superstring field theory
and avoid violating the above three arguments for the following three
reasons: 1) The standard R-R vertex operator, 
$c_L e^{-\phi_L/2} \Sigma_{L\a}~
c_R e^{-\phi_R/2} \Sigma_{R}^\b$, is actually the vertex operator for the
auxiliary field $F_{(0)\b}^\a$ and not for the gauge field. Only on-shell,
this auxiliary field is the field-strength for the gauge field;
2) Because $V_{(0)}^{\a\b}$ is BRST-equivalent with vertex operators of
non-vanishing $J_L-J_R$ ghost-number, it appears BRST-trivial in
the standard GS formalism where ghosts are not yet understood. The
confusion about ghosts in the standard GS formalism is probably related
to the absence of a Fradkin-Tseytlin term in the standard GS sigma model
since the zero-momentum
``ghost'' dilaton is also described by a vertex operator of
non-vanishing $J_L-J_R$ ghost number; 3) Since all perturbative superstring
states can be described by vertex operators of finite $J_L-J_R$ ghost
number, they do not couple to the zero-momentum
R-R fields. However,
$D$-branes are described by boundary states which contain all possible
$J_L-J_R$ ghost numbers, allowing them to couple to zero-momentum R-R 
fields.\ref\Dbrane{J. Polchinski and Y. Cai, Nucl. Phys. B296 (1988) 91\semi
C. Callan, C. Lovelace, C.R. Nappi and S.A. Yost, Nucl. Phys. B308
(1988) 221\semi
S.A. Yost, Nucl. Phys. B321 (1989) 629.}\sftwo

The next step in the analysis is to write the zero-momentum vertex operators,
$V_{(n)}^{\a\b}$ and 
$V_{(n)\a\b}$, in integrated form. RNS unintegrated vertex operators
carry negative picture (e.g., $V=c_L e^{-\phi_L} \psi_L^\mu
c_R e^{-\phi_R} \psi_R^\mu$ for the graviton), so one first needs to perform
a picture-raising operation before anti-commuting with
$\int dz_L b_L$ and
$\int dz_R b_R$. 
Because the rule for picture-raising
comes from a complicated closed superstring
field theory argument, its justification will be left for a separate paper.
In this paper, it will be enough to know that the ``picture-raised''
version of $V_{(n)}^{\a\b}$ is given by multiplying 
$V_{(n)}^{\a\b}$ with $(\p_L c_L-\p_R c_R)$, then
multiplying with $\xi_L$ and taking the second-order pole with
$b_R\xi_R$, and finally
commuting with $Q_R$ and with $Q_L$. ($c_L^0-c_R^0$ removes
the $b_L^0-b_R^0$ constraint, $Q_L\xi_L^0$ is the picture-raising operator
for fields, and $Q_R (b_R\xi_R)^0$ is the picture-raising operator
for anti-fields.) For the 
usual states of ghost-number (1,1), this is equivalent to 
multiplying with $\xi_L \xi_R$ and then commuting with
$Q_L$ and $Q_R$.

For $V_{(0)}^{\a\b}$ of \vdef, multiplication with $\xi_L\xi_R$ gives
\eqn\mult{
\xi_L\xi_R V_{(0)}^{\a\b}= c_L \xi_L e^{-3\phi_L/2} \Sigma_{L}^{\a} ~
c_R \xi_R e^{-\phi_R/2} \Sigma_{R}^{\b}={1\over{160}}
\tb_L^\a (\t_R\G^\mu\tb_R)
\G_\mu^{\b\g}\t_{R\g}}
where $\t^\a$ and $\tb^\a$ are defined in \thetadef.\foot 
{Upon compactification to
D=4, this vertex operator becomes $\tb_L^\ad (\tb_R^\gd \tb_{R\gd}) 
\t_R^{\beta}$, which is the vertex operator for the graviphoton in
the N=2 D=4 supergravity multiplet.\calabi\ref\eff{N. Berkovits and
W. Siegel, Nucl. Phys. B462 (1996) 213.}} Since $\{Q_L~,~\tb_L^\a\}$=0,
there is no integrated vertex operator associated with $V_{(0)}^{\a\b}$.
For this reason, there is no candidate for a zero-momentum R-R vertex operator 
in the standard GS sigma model.

However, one can also ask what is the integrated form of the 
zero-momentum R-R vertex operators $V_{(n)}^{\a\b}$ for $n>0$.
In fact, just as the ``ghost'' dilaton is necessary for preserving manifest
reparameterization invariance, the ``ghost'' version of the R-R field
is necessary for preserving manifest spacetime supersymmetry. This is
easiest to see in D=4 Type II superspace effective actions\eff where manifest
N=2 D=4 supersymmetry requires the graviphoton field to appear
both in the supergravity multiplet (whose vertex operator is $V_{(0)}^{\a\bd}$)
and in the vector compensator multiplet (whose vertex operator is
$V_{(1)}^{\a\bd}$). In other words, fixing $D_{(1)\a\b}=
E_{(1)}^{\a\b}
=0$ gauge-fixes
part of the super-reparameterization invariances, so one needs to keep
$D_{(0)\a\b},
E_{(0)}^{\a\b}$
and
$D_{(1)\a\b},
E_{(1)}^{\a\b} $ in the action if one wants to preserve manifest
spacetime supersymmetry. It is unclear at the moment if one can gauge
away the $D_{(n)\a\b}$ and $E_{(n)}^{\a\b}$ fields for
$n>1$ without
breaking manifest spacetime supersymmetry.

After performing the picture-raising operation and
anti-commuting with $\int dz_L b_L$ and $\int dz_R b_R$,
the integrated form of 
$V_{(1)}^{\a\b}$ is
$$\{\int dz_L b_L~,~c_L\xi_L e^{-3\phi_L/2} \Sigma_L^\a\}~
\{\int dz_R b_R~, ~[Q_R~,~(b_R c_R +\p_R\eta_R)e^{3\phi_R/2}\Sigma_R^\b 
]~\}$$
$$=
\int dz_L \xi_L e^{-3\phi_L/2} \Sigma_L^\a~
\{Q_R~,~ -\int dz_R b_R e^{3\phi_R/2}\Sigma_R^\b \}$$
$$=
\int dz_L \xi_L e^{-3\phi_L/2} \Sigma_L^\a~
\{Q_R~,~\int dz_R  e^{\phi_R/2}\Sigma_{R\g}\p_R x^\mu \G_\mu^{\b\g} \}$$
$$=
\int d^2 z ~\lb_L^\a~
\{Q_R~,~ \t_{R\g}\p_R x^\mu \G_\mu^{\b\g} \}$$
\eqn\final{=
\int d^2 z \lb_L^\a~
\l_{R\g}\p_R x^\mu \G_\mu^{\b\g} }
plus terms which are independent of $x^\mu$, where it was used that
$Q_R$ anti-commutes with $\bar q^\b_R$ of \suba.

This integrated vertex operator is BRST-trivial
since it can be written as the anti-commutator of $Q_L+Q_R$ with
$\int d^2 z ~\lb_L^\a 
\t_{R\g}\p_R x^\mu \G_\mu^{\b\g}$. (Note that $[Q_L,\int dz_L \lb_L^\a]$=
$\int dz_L \p_L\tb_L^\a$=0.) Nevertheless, it is not identically zero,
which
means there should be a term in the 2D sigma model of the form
\eqn\form{N \int d^2 z ( D_{(1)\a\b}~ \lb_L^\a
\l_{R\g}\p_R x^\mu \G_\mu^{\b\g} 
+ E_{(1)}^{\a\b} ~\l_L^{\g}\p_L x^\mu \G_{\mu\,\a\g}\lb_{R\b})}
where $N$ is an as yet undetermined normalization factor.

As discussed in \sft\sftwo and as implied by the equations of motion for
$F_{(0)\b}^\a$ in \motion, the graviphoton gauge field $A_\mu$ is given by
$A_\mu=D_{\a\b}\G_\mu^{\a\b} +E^{\a\b}\G_{\mu\,\a\b}$.
So its 
vertex operator has a term
$N\int d^2 z$
$( \l^A \G^{11}_{AB} \lb^B )\pt x^\mu$,\foot{The
vertex operator for $A_\mu$ also has a
term proportional to 
$\int d^2 z $
$(\l^A \G^{11}\G^{\mu\nu}_{AB} \lb^B)$
$ \pt x_\nu$, 
but this term
does not seem to have a 
higher-dimensional 
interpretation.} 
suggesting that
$N\oint d\sigma \l^A \G^{11}_{AB} \lb^B$ can be associated with the
zero-brane R-R charge. This agrees with the proposal of \rnsd
if $N={1\over{32}}.$ 

Furthermore, the R-R three-form gauge field $A_{\mu\nu\rho}$ 
is given by 
$A_{\mu\nu\rho}$
$=D_{\a\b}\G_{\mu\nu\rho}^{\a\b} +$
$
E^{\a\b}\G_{\mu\nu\rho\,\a\b}$, so its vertex operator contains the term
${1\over{32}}\int d^2 z $
$(\l^A \G^{12\,[\mu\nu}_{AB} \lb^B) $
$\pt x^{\rho]}$, suggesting
that the
two-brane R-R charge can be identified with
$\oint d\sigma$
$ \l^A \G^{12\,\mu\nu}_{AB} \lb^B$. This fits beautifully
with the D=11 interpretation since the NS-NS two-form gauge field 
$B_{\mu\nu}$ has the vertex operator
$\int d^2 z \p_\sigma x^{[\mu}\p_\tau x^{\nu]}$=
${1\over{32}}
\int d^2 z (\l^A \G^{12\,11\,[\mu}_{AB} \lb^B ) \pt x^{\nu]}$, and the
one-brane NS-NS charge is
$ {1\over{32}}\oint 
d\sigma \p_\sigma x^\mu$=
$\oint d\sigma \l^A \G^{12\,11\,\mu}_{AB} \lb^B$.
($\t^\a$ dependence is being ignored in this
comparison.)

Also, one can identify the 
four-brane 
R-R charge with 
$\oint d\sigma$
$ \l^A \G^{11\,\mu\nu\rho\kappa}_{AB} \lb^B$. 
Note that all of these R-R charges are 
BRST-trivial, but
are not identically zero.

\subsec{2D sigma model and $e^{\phi}$ dependence}

Actually, in the 2D sigma model, ${1\over{32}}
\int d^2 z (\l \G^{11}\lb) \pt x^\mu$
must couple to $e^{\phi} A_\mu$, rather than simply $A_\mu$. ($<e^{\phi}>$
is the string coupling constant.) The reason is that the tree-level
effective action for $A_\mu$ is ${1\over 4}\int d^{10}x 
\p_{[\mu} A_{\nu]}\p^{[\mu} A^{\nu]}$ with no $e^{-2\phi}$ factor, so
$e^{\phi}$ needs to appear with $A^\mu$ in order to cancel the
overall $e^{-2\phi}$ which comes at string tree-level from the
Fradkin-Tseytlin term
$\int d^2 z \phi R$. Note that this $e^{\phi}$
dependence is invisible when expanding the 2D sigma model to 
first order around a flat D=11 supergravity background (since
$e^{\phi}=(g_{11\,11})^{3/4}=1$ in a flat D=11 background).

Since the zero-brane charge (or equivalently $P_{11}$) is given by
$\d {\cal S}/\d\Lambda$ 
where ${\cal S}$ is the sigma model action and
$\d A_\mu=\p_\mu\Lambda$
\ref\az
{J.A. de Azcarraga, J.P. Gauntlett, J.M. Izquierdo and P.K. Townsend, 
Phys. Rev. Lett. 63 (1989) 2443.},
this suggests that $P_{11}={1\over{32}} 
e^{\phi}(\l\G^{11}\lb)$.

\newsec {Justification based on $D_0$-brane analysis}

One of the most compelling arguments for an eleven-dimensional origin
of superstring theory is that $D_0$-branes can be interpreted as
Kaluza-Klein states coming from compactification 
on a circle 
of D=11 supergravity states.\dual\witten\polch
As the radius of the circle goes to zero
(which corresponds to the string coupling constant going to zero),
these Kaluza-Klein states become infinitely massive in the D=10 metric.
However, in terms of
the eleven-dimensional metric, the $D_0$-branes are massless states.
It will now be shown that identification of $P_{11}$ with
${1\over{32}}
e^{\phi}(\l\G^{11}\lb)$ agrees with this picture.

The first step is to define the $D$-brane boundary conditions for the
twistor variables. This is easy since the spinor superfields
$\T_L^\a$, $\Tb_L^\a$ and $\T_{R\a}$, $\Tb_{R\a}$ satisfy the same
$D$-brane boundary conditions as their lowest component $\t$ variable.
Therefore, at the end of an open string with Neumann boundary conditions
in directions $0 ... P$ and Dirichlet boundary conditions in directions
$(P+1) ... 9$, the boundary conditions
on $\t^\a$ and $\l^\a$ are given by
\eqn\bound{\T_L^\a= (\G^0 ... \G^P)^{\a\b} \T_{R\b},\quad
\Tb_L^\a= (\G^0 ... \G^P)^{\a\b} \Tb_{R\b}}
where $P$ is assumed to be even.

It is easy to check this implies that
$$\p_L x_L^\mu ={1\over{16}}(\l_L \G^\mu\lb_L +i(\p_L\t_L)\G^\mu\tb_L)
=\pm{1\over{16}}(\l_R \G^\mu\lb_R +i(\p_R\t_R)\G^\mu\tb_R)=\pm \p_R x_R^\mu$$
where the plus sign is if $\mu \leq P$ and the minus sign is if
$\mu >P$.

So if the end of the open string lies on a $D_0$ brane at rest, 
\eqn\argy{P_{11}={1\over{32}}
e^{\phi}(\l_L^\a\lb_{R\a}+\l_{R\a}\lb_L^\a)
={1\over{16}} e^{\phi}(\l_R\G^{0}\lb_R)=e^{\phi} P_0.}
Also, $P_\mu=\pt x_\mu =0$ for $\mu=1 ... 9.$

Therefore, the $($mass$)^2$ of the $D_0$-brane in the eleven-dimensional
metric $G^{MN}$ is 
\eqn\mass{M^2= G^{MN} P_M P_N = G^{00} (P_0)^2 +G^{11\, 11} (P_{11})^2}
$$=e^{2\phi/3} (P_0)^2 - e^{-4\phi/3} (e^\phi P_0)^2 =0$$
where the ten-dimensional metric has been assumed to be flat, so
using the conventions of \witten, $G^{\mu\nu}=e^{2\phi/3}\eta^{\mu\nu}$
and $G^{11\, 11}= - e^{-4\phi/3}$. Furthermore, the $($mass$)^2$
in the ten-dimensional metric, $(P_0)^2$, can be computed using
standard $D$-brane techniques\polch, and diverges like
$e^{-2\phi}$ as the string coupling constant goes to zero.

So the identification of $P_{11}$ with 
${1\over{32}} 
e^{\phi}(\l_L^\a\lb_{R\a}+\l_{R\a}\lb_L^\a)$ is supported
by the Kaluza-Klein picture of the $D_0$-brane.

A further check on this identification of $P_{11}$ comes from
the N=2 D=10 SUSY algebra of the superstring. As discussed in
\ref\rrc{N. Berkovits, `Ramond-Ramond Central Charges in the
Supersymmetry Algebra of the Superstring', IFT-P.038/97, hep-th
9706024.}, the N=2 D=10 SUSY algebra contains Ramond-Ramond
central charge terms of the form 
\eqn\rrcent{\{q^\a_L, q_{R\b}\} = 
C_{(0)\b}^\a }
where $C_{(0)\a}^{\a}$ is the zero-brane Ramond-Ramond central
charge which will be identified with 
$\oint d\s(\l_L^\a\lb_{R\a}+\l_{R\a}\lb_L^\a)$.
This can be compared with the N=1 D=11 SUSY algebra, 
\eqn\elev{\{\hat q^A, \hat q^B\}= \G_m^{AB} E^{m M} P_M}
where $A=1$ to 32 are SO(10,1) spinor indices,
$m=0, ..., 9, 11$ are flat vector indices,
$M=0, ..., 9, 11$ are curved vector indices, $\hat q^A$ are
the N=1 D=11 SUSY generators, and $E^{mM}$ is the
D=11 vierbein.
When the D=10 metric is flat, $E^{mM}= e^{{1\over 3}
\phi}\d^{mM}$ for $m$=0 to 9
and $E^{11 ~M}= e^{-{2\over 3}\phi}\d^{11~ M}$.\witten
So if $P_M$ is the ten-dimensional momentum for $M=0$ to 9,
then $\hat q^A$ should be identified with $e^{{1\over 6}\phi} q_L^\a$
for $A$=1 to 16 and with 
$e^{{1\over 6}\phi} q_{R\a}$ for $A$=17 to 32. Comparing \elev
with \rrcent, this implies that 
$P_{11}={1\over{32}} e^{\phi} C_{(0)\a}^{\a}.$  

\newsec{Extra dimensions in the Type IIB superstring}

All of the previous constructions 
for the Type IIA superstring can be repeated for the Type IIB
superstring, however because $\T_L^\a$,$\Tb_L^\a$ and
$\T_R^\a$,$\Tb_R^\a$ now have the same SO(9,1) chirality, they
cannot be combined into 32-component SO(10,1) spinor superfields.

But they can be combined into $16\times 2$-component SO(9,1)$\times$SO(2,1)
spinor superfields, $\T_b^\a$ and $\Tb_b^\a$, where $b=1$ for the left-moving
superfield and $b=2$ for the right-moving superfield. In this notation,
the momentum and one-brane NS-NS charge can be written as 
\eqn\zero{{1\over{32}}
\oint d\sigma\l_b^\a \tau^{bc}_0 \G^\mu_{\a\b}\lb_c^\b,\quad
{1\over{32}}
\oint d\sigma\l_b^\a \tau^{bc}_1 \G^\mu_{\a\b}\lb_c^\b,}
where $\tau_q^{bc}$ are $2\times 2$ SO(2,1) $\G_q$ matrices which
are related to the usual SO(2,1) $\g_q$ matrices by multiplication
with $\g_0$ (i.e. $\tau_0^{bc}=\d^{bc}$,
$\tau_1^{bc}=\sigma_3^{bc}$,
$\tau_2^{bc}=\sigma_1^{bc}$).

This suggests that the one-brane 
R-R charge should be identified with
${1\over{32}}$
$\oint d\sigma$
$ \l_b \tau_2^{bc} \G^\mu \lb_c$,
and such an identification is supported by 
arguments based on 
zero-momentum Type IIB R-R vertex operators which are similar to
those of subsection (3.2). Furthermore, 
the three-brane R-R charge
is naturally identified
with ${1\over{32}}
\oint d\sigma \l_b \e^{bc} \G^{\mu\nu\rho} \lb_c$, and
the five-brane R-R charge
with ${1\over{32}}
\oint d\sigma \l_b \tau_2^{bc} \G^{\mu\nu\rho\kappa\phi} \lb_c$.

\newsec{Conclusions}

There is increasing evidence that superstring theory is part of an
eleven-dimensional structure. In this paper, it was proposed that
superstring theory itself can be used to understand the eleventh dimension.
If this proposal turns out to be correct, one should be able to understand
$M$-theory compactifications using superstring language. Since the extra
dimension is built out of RNS matter and ghost fields,
perhaps one will need to consider superstring
compactifications which treat the RNS matter
and ghost degrees of freedom on an equal footing. Note
that using the N=0 $\to$ N=1 embedding of reference \ref\vafa
{N. Berkovits and C. Vafa, Mod. Phys. Lett. A9 (1994) 653.},
heterotic and Type II superstring backgrounds can be treated
equivalently if the distinction between matter and ghost fields
is removed. This suggests that invariances
which transform the RNS matter
and ghost fields into each other might be related to
superstring dualities.

An alternative proposal for understanding the eleven-dimensional structure
is M(atrix) theory 
\ref\M{T. Banks, W. Fischler, S.H. Shenker, L. Susskind,
Phys. Rev. D55 (1997) 5112.},
which is closely related to a light-cone GS approach
in eleven dimensions.
It is interesting to note that covariantization of the light-cone GS
superstring in ten dimensions 
was one of the main motivations for studying the twistor-GS
formalism.\het Perhaps the introduction of twistor variables into M(atrix)
theory will allow a more SO(10,1)-covariant description of the theory.

Another proposal for understanding the eleventh dimension uses an N=(2,1)
heterotic string to generate target spaces which are either the
supermembrane worldvolume or the superstring worldsheet. \ref\mart{
D. Kutasov and E. Martinec, Nucl. Phys. B477 (1996) 652\semi
D. Kutasov and E. Martinec, ``M-Branes and N=2 Strings, hep-th 9612102.} 
This formalism shares with the twistor-GS superstring the property of
having
N=2 worldsheet supersymmetry.\foot{This fact was recently used by DeBoer and
Skenderis\ref\sk{J. deBoer and K. Skenderis, ``Self-Dual Supergravity
from N=2 Strings'', hep-th 9704040.} to construct a hybrid twistor-heterotic
string theory which describes self-dual supergravity.} 
In the N=(2,1) heterotic approach, the difference between the
supermembrane and superstring comes from the choice of a null
superconformal constraint. If this null constraint could be interpreted
as a gauge-fixing condition, it would mean that the $M$-theory
variables and superstring variables were related by a field redefinition
which connects the two different gauge choices.
This sounds similar to the proposal of this paper, however it is
difficult to verify since only
the static-gauge $M$-theory and superstring variables are easily obtainable
in the N=(2,1) heterotic approach.

{\bf Acknowledgements:} I would like to thank
S. Ramgoolam, W. Siegel and B. Zwiebach
for useful conversations. I would also like thank
the IAS at Princeton and the ITP of SUNY at Stony Brook
where part of this work was done. This
work was financially supported by 
FAPESP grant number 96/05524-0.

\listrefs
\end